\documentstyle[12pt]{article}
\begin{document}

\begin{titlepage}
\vspace{7mm}
\centerline{\LARGE P.G.Grinevich and S.P.Novikov}
\vspace{7mm}
\centerline{\LARGE String equation--2. Physical solution.}
\vspace{5mm}
\centerline{\Large Table of Contents}
\vspace{5mm}
{\large 1.Introduction}

{\large 2.General identities}

{\large 3.String equation and deformations of the Riemann surfaces}

{\large 4.Semiclassical approximation and foliations of the Riemann
surfaces}

{\large 5.Special Semiclassics for the
Lax pairs in the case of Physical solution}

{\large 6.Proof of the main theorems of the p.5}

{\large References}
\end{titlepage}
\vspace{7mm}
{\bf 1.Introduction}

\vspace{5mm}

This paper is a continuation of \cite{nov1}\footnote{Russian version of this
paper was published in the Journal 'Algebra and Analysis'
(St--Petersburg, Russian Academy of Sciences, 1994), v.6, No 3, pp. 118-140,
dedicated to the 60--ieth birthday of L.D.Faddeev}.

{\bf String Equation
is by definition the equation $[L,A]=1$ for the coefficients
of two linear ordinary differential operators $L$ and $A$.}

This terminology appeared after the papers \cite{m-g,b-k,d-s} in 1989-90.
For the so-called ''double-scaling limit of 1-matrix models''
we always have $L=-\partial_x^2+u(x)$, $A$ is some operator of
the odd order $2k+1$. For $k=0$ we have trivial answer
$u=x, \; A=\partial$.

First nontrivial case is $k=1$.
We have here Gardner--Green--Kruscal--Miura--Lax operator
\[A=-4\partial^3+6u\partial+3u_x\]

and the following differential equation (string equation)
for $u(x)$:
\[u_{xxx}-6uu_x=1\]

Last equation after one integration exactly coinsides with
the well--known Painlev\'{e} equation of the first type or
simply {\bf P-1 equation}:
\begin{equation}
u_{xx}-3u^2=x \label{eq:p1}
\end{equation}

We are looking for real solutions on the $x$--line or at least
on the halfline $-x\rightarrow\infty$.
Even more: the solutions which may
be important for the Quantum String Theory should have
asymptotics
\begin{equation}
u(x)\sim +\sqrt{-x/3},\; x\rightarrow -\infty \label{eq:psol}
\end{equation}

For the mathematical foundation of the 'double--scaling limit'
 and of the selection of Physical Solution of P-1 equation
  from the finite matrix models see \cite{fik}.

P-1 equation has two types of solutions in the class
of formal series:
\begin{equation}
u_{\pm}(x)=\pm\sqrt{-x/3}
(1+\sum_{i=1}^{\infty}
a^{\pm}_i \tau^{-2i}),\; \tau=(-x)^{5/4}  \label{eq:fsol}
\end{equation}

{\bf It was shown by Boutroux \cite{bou} that there exists an unique
 exact solutions of P-1, for
which the formal series $u_-(x)$ is an asymptotic one.}

For other solutions such that $u\sim -\sqrt{-x/3}$
even the second coefficient $a^-_1$ has nothing in common
  with the real asymtotic behavior.

Situation is different for the physical
solutions $u_+(x)$:

 {\bf There exists an  1-parametric 'separatrix' family
  of  solutions of the P-1 equation for which  formal
   solution $u_+(x)$ is an asymptotic series.}

More detailed information about these families can be found in the
work \cite{h-s}.

   In the papers \cite{nov1,g-m} a 'Plank parameter'
   has been introduced in the theory of P-1 equation.
   For studying its asymptotics it is good to write it
   in the form
   \begin{equation}
   [L,A]=\epsilon \cdot 1,\;u_{xx}-3u^2=\epsilon \cdot x
   \label{eq:p11}
   \end{equation}

Two asymptotic methods were developed in \cite{nov1}
for the equation above:
{1.\bf Nonlinear semiclassics (or averaging or Bogolyubov--
Whitham method)}--
see Appendix in \cite{nov1} written by Dubrovin
and Novikov. This method has been
improved essentially in \cite{kri1}. However it gives
some good information about some nonphysical solutions
 only. For example,
 Dubrovin and Novikov in \cite{nov1}, Appendix,
 studied by this method some solutions with infinite number of poles
 on the important real halfaxis
 \(-x\rightarrow\infty\);
Krichever investigated nonphysical solutions of the type
$u\sim -\sqrt{-x/3}$ in \cite{kri1}.

We don't see any way to apply this technic to the studying of the
physical solutions for $k=1$. The nonphysical solutions  mentioned
 above
 have been studied also earlier in the work \cite{kop} and
 in the later papers of the St-Petersburg group
 (see the survey \cite{kit}; they were known earlier in
 the classical theory of P-1 equation).

 {\bf 2.Linear semiclassics for the Lax pair.}

 The idea of the 'Isomonodromic' method has been invented  in the
 work \cite{f-n}. It is based on the semiclassics.
 It has been applied to the different
 Painlev\'{e} types in \cite{m-j}. P-1 equation
 was studied for example in \cite{kop,g-m,kk,kit}. Some results
 concerning the behavior of physical solution along
 the special lines in complex $x$-space and monodromy
 ('Stokes') coefficients were obtained.

 In the work
 \cite{nov1} completely different semiclassical ideas
 were developed. In particularly some strange beautiful identities
 connecting the theory of P-1 with families of elliptic
 curves were found. These ideas and their developement may be
 found in the paragraphs 2--4 of the present work.

 Some very specific features
 of the "Isomonodromic' method available for the Physical
 solution only
 will be developed it the present work (see the paragraphs 5 and 6).
 The comparison of our technic with St-Petersburgs's and Fokas group
  may be found in the Remark after the proof of the Corollary 1
  in the paragraph 5.

   {\bf  By the conjecture of Novikov, these special 'Physical'
    solutions of the equations like $[L,A]=1$, which are
    analytically exceptional, probably have much
    stronger 'hidden symmetry' than other solutions
    of the same equations and are much more
    'exactly integrable' than others. However, this problem
    is not yet solved.}

    Our paper is written for the case $k=1$ only. However there are
    no difficulties to develope the same staff for any $k$.
    For even values of $k=2n$ the physical solutions can be described
    by the method 1 above, like in \cite{kri1}, but the case $k=1$
    is more complicated.

     \pagebreak

\vspace{7mm}

    {\bf 2.General Identities.}

    \vspace{5mm}

    Let us construct a 'Zero--Curvature Representation' for the P-1
     equation, starting from the zero-curvature representation
     of the KdV system found in the very first paper of Novikov
     on the periodic problem \cite{nov2,dmn}
     (it was more convenient for the
      studying of the so-called 'finite-gap solutions'
      than the ordinary
      Lax representation $L_t=[L,A]$). KdV is equivalent
      to the compatibility condition for the pair
      of the linear systems below:
      \begin{eqnarray}\Psi_t=\Lambda\Psi,\; \Psi_x=Q\Psi\\
      \Lambda_x-Q_t=[Q,\Lambda ] \end{eqnarray}
      Here we have \[\Lambda=\left( \begin{array}{cc} a&,b\\c&,-a
      \end{array}\right),\;Q=\left( \begin{array}{cc}
      0&,1\\u-\lambda&,0\end{array}\right)\]
      and
      \[a=-u_x,\;b=2u+4\lambda,\;c=-4\lambda^2+2\lambda u+
      2u^2-u_{xx}\]
      You may replace here $u_{xx}$ by $\epsilon\cdot x+3u^2$

     {\bf Remark.}In general, for any $k$
     we have analogous representation with the same matrix $Q$,
     $b$ is a polinomial of the order $k$ in $\lambda$

\newtheorem{prop}{Proposition}
\begin{prop}
Replacing the derivative $\partial_t$ by $\epsilon\cdot
\partial_{\lambda}$ we get a 'Zero-Curvature' representation
for  P-1 equation  \ref{eq:p11}
\begin{eqnarray}\epsilon\Psi_{\lambda}=\Lambda\Psi,\;
\Psi_x=Q\Psi\\ \Lambda_x-\epsilon \cdot Q_{\lambda}=-[\Lambda,Q]
\end{eqnarray}
\end{prop}

For the proof, it is easy for example to check this
by elementary calculation. In fact,before 1990 people studied
P-1 system using some different representation
(see \cite{m-j,kop,kit}). This one directly follows
from the Heisenberg type relation $[L,A]=1$, as it was
pointed out in \cite{nov1}. It looks more natural.

Let \(R(\lambda)=-\det \Lambda=a^2+bc=-16\lambda^3-4C\lambda-D\).
\begin{prop}
The following general identities are true:
\begin{eqnarray}dR/dx&=-\epsilon\cdot b,\;db/dx=-2a\\
dC/dx&=\epsilon,\;dD/dx=2\epsilon\cdot u \\
C&=u_{xx}-3u^2,\;D=-4u^3-u_x^2+2uu_{xx} \\
u_x&=\sqrt{R(\lambda)}|_{\lambda=-u/2}
\end{eqnarray}
\end{prop}

This statement also might be checked for $k=1$ by the elementary
calculation. Its analog follows easily from the
'Zero-Curvature Representation' for any $k$. It is obvious
that all coefficients of the Riemann surface, who were
constants for $\epsilon=0$ will have $x$-derivatives
 proportional to $\epsilon$.

Last equation demonstrates very nice interpretation of P-1
as a special deformation of the elliptic Riemann surface
$\Gamma(x)$:
\[y^2=R(\lambda)\]
(Here $x$ plays a role of the parameter). We shall discuss
this subject in the second paragrath.

Let us itroduce a $Z_+$-graded commutative
and assotiative differential ring $A_{\epsilon}$
over $Z[1/2]$
 containing the symbols \(u,\partial,\epsilon\)
such that:
\begin{eqnarray}A_{\epsilon}&=\sum_{i\geq 0} A_i, \;
A_0=Z,\; A_1=0\\
u&\in A_2, \; u_x=\partial u\in A_3,\; u_{xx}=\partial^2u\in
 A_4,\ldots,\epsilon\in A_5\\
 u_{xxx}&-6uu_x=\epsilon
 \end{eqnarray}

\newtheorem{theorem}{Theorem}
\begin{theorem}
Consider the graded rings $A_{\epsilon}$
 and a trivial $\epsilon$-extension of the
 ring of elliptic functions
holomorphic outside of the point $a=0$  on the algebraic curve
$\Gamma$ (the same as above)
\[\tilde{y}^2=\tilde{R}(\mu)=4\mu^3-g_2\mu-g_3,\;\mu=-\lambda,\;
\tilde{y}=y/2,\;g_2=-C=-\epsilon\cdot x,\;g_3=D/4\]
and a substitution, which preserves
 the grading by definition:
\[u(x)=2\wp(a)\]
(the grading of $\wp^{(n)}(a)$ we take as $2+n$).
The following formulas are true:
\begin{eqnarray}
u=2\wp(a),\; u_x=2\wp '(a)=2\partial_w\wp(w+a)|_{w=0} \nonumber \\
 u_{xx}=2\wp ''(a)=\partial_w^2\wp(w+a)|_{w=0} \nonumber \\
u_{xxx}=2\wp '''(a)+\epsilon \nonumber \\
u^{(4)}_{x\ldots x}=2\wp^{(4)}(a),\; u^{(5)}_{x\ldots x}
=2\wp^{(5)}(a)+6u\epsilon \nonumber \\
 u^{(6)}_{x\ldots x}=2\wp^{(6)}(a)+24u_x\epsilon  \nonumber  \\
 u^{(7)}_{x\ldots x}=
 2\wp^{(7)}(a)+(60u_{xx}+36u^2)\epsilon \nonumber \\
 u^{(8)}_{x\ldots x}=2\wp^{(8)}(a)+(36\cdot 30 uu_x)\epsilon+
 60\epsilon^2  \nonumber \\
 \ldots \nonumber\\
u^{(n)}_{x\ldots x}=2\wp^{(n)}(a)+
\sum_{i\geq 1}P_{i,n}(u,u_x,u_{xx})
\epsilon^i, P_i\in A_{n-5i+2}
\end{eqnarray}
 The polinomials $P_{i,n}$ have constant integral
 coefficients.  They
 depend only on the variables $u,u_x,u_{xx}$, which
 may be replaced by \(\wp,\wp ',\wp ''\).
  \end{theorem}

  {\bf Remark.} After the identification of $\epsilon$
  with scalars we shall have a ring, graded modulo 5.
  Identifying the symbols $g_2,g_3$ with the scalars
  we shall destroy the grading, but the formulas
  above don't contain these symbols.

   The proof of this theorem is based on the
     equation (which is already established)
   \(u_x=2\sqrt{\tilde{R}(u/2)}\)
   and on the comparison of ( \ref{eq:p11}) with
   standart equations for the elliptic functions
   (see \cite{b-e}):
   \begin{eqnarray*}(\wp ')^2=4(\wp)^3-g_2\wp-g_3 \\
   g_2 '=g_3 '=\epsilon '=0\\
   \wp ''=6(\wp)^2-g_2/2
   \end{eqnarray*}

  The equality $u_{xx}=2\wp ''(a)$ follows from the relation
  $u_{xx}/2=6(u/2)^2-g_2/2$, the same as for $\wp ''$.
  Differentiating it and using the relation
  $u_{xxx}=6uu_x+\epsilon$, we obtain the next relation
  for $u_{xxx}$ and so on. Each time when $u_{xx}$ appears,
  its derivative in $x$ will contain $\epsilon$. The total
   sum of
    terms in the final sum which don't
    contain $\epsilon$ at all will exactly equal to
    the formula for $\wp^{(n)}$, because they are constructed
    exactly by the same rule. The theorem therefore is proven.
    \pagebreak
    \vspace{7mm}

    {\bf 3.String equation and deformations of the Riemann Surfaces}

    \vspace{5mm}

    Consider now the Zero-Curvature equation for P-1 (with
     $\epsilon$) and especially the first linear system
     in the variable $\lambda$. It contains a parameter $\epsilon$
     as a Plank constant. It is natural therefore
      to develope a semiclassical approach to the studying
      of the $\Psi(\lambda)$, because the
      $\lambda$-dependence of the coefficients is already known
      (the corresponding matrix $\Lambda$ is polinomial
      and traceless, it does not depend on $\epsilon$).

      {\bf The first principle of Semiclassical Approach
      is that it should
      be applied to systems which are diagonal neglecting
       the terms of the order $\epsilon$ in the righ-hand
       part}.

       Therefore we should make a corresponding transformation.
       Consider a matrix $U(\lambda)$, such that
       \begin{eqnarray}
       U^{-1}\Lambda U=\left( \begin{array}{cc}
       \sqrt{R}&0 \\
       0&-\sqrt{R}
       \end{array} \right)=\Lambda ' \nonumber \\
       R=-\det \Lambda=a^2+bc \nonumber \\
       U=\left( \begin{array}{cc}
       1&1 \\ \chi_-&\chi_+
       \end{array} \right) \nonumber \\
       \chi_{\pm}=\frac{-a\mp\sqrt{R}}{b}
       \end{eqnarray}

     After the substitution  \[ \Psi=U\tilde{\Psi} \]
     we are coming to the new system  through the
     Gauge Transformation
     \begin{equation}
     \epsilon \tilde{\Psi}_{\lambda}= \tilde{\Lambda}\tilde{\Psi}
     =(\Lambda '-\epsilon \cdot U^{-1}U_{\lambda})\tilde{\Psi}
     \label{eq:diag}
     \end{equation}

     The last system is diagonal modulo $\epsilon$, but
     {\bf It is defined on the Riemann surface}
     \[ \Gamma :
     y^2=R(\lambda)\]

     not on the complex plane.

     After that we may write a formal semiclassical
     solution:
     \begin{prop}
     There exists a formal solution
     \begin{equation}
     \tilde{\Psi}_{sc}=(1+\sum_{i\geq 1} A_i \epsilon^i)
     e^{ \{ \frac{1}{\epsilon} B_{-1}+B_0+\sum_{i\geq 1}B_i
     \cdot \epsilon^i\}}
     \label{eq:sc}
      \end{equation}
     such that all  matrices $A_i$ have diagonal elements
    equal to zero, all matrices $B_i$ are diagonal
    and may be found in the form:
    \begin{eqnarray} B_{-1,\lambda}=\sqrt{R}
    \left( \begin{array}{cc}1&0 \\0&-1
    \end{array}
    \right) \nonumber \\
    B_{0,\lambda}=\left( \frac{a}{b} \right)_{\lambda}
    \frac{b}{2\sqrt{R}} \left( \begin{array}{cc}
    1&0 \\ 0&-1 \end{array} \right)-\left( \frac{R_{\lambda}}{4R}
    -\frac{b_{\lambda}}{2b} \right )\cdot 1
    \end{eqnarray}
    All functions ---matrix elements of
    $A_i$ are algebraic on $\Gamma$, all differential
    forms---matrix elements  of $dB_i(\lambda)$ are abelian differentials
    on $\Gamma$.
    \end{prop}

   {\bf Remark 1.}There exist a formal solution
   of this form which satisfies
    to the both linear equation in the Zero-Curvature
    representation for the equation ~\ref{eq:p11}.
    We shall discuss this later.

    {\bf Remark 2.} It is easy to write algebraic formulas
    for all $A_i$ and \( B_{i,\lambda} \), like in \cite{nov1}.
    The last proposition in fact is extracted from this work.

   The proof of this proposition may be obtained by the direct
   substitution in the equation.  All formulas for  matrix
   elements were written in \cite{nov1}.

   There is a very interesting analogy between semiclassics
    for this  system and the so-called 'Baker--Akhiezer'
    function:
      \begin{prop}
      The semiclassical solution above is essentially scalar
       function on the  Riemann surface $\hat{\Gamma}$ which is a covering
       over the surface $\Gamma$ with some branching point.
        It means
        exactly that the permutation of the matrix indices
        1 and 2 is equivalent to the permutation $\kappa$ of
       of   sheets of the surface $\Gamma$. In particularly, we have
       \begin{eqnarray}
       \kappa^{\ast}db_{11}(\lambda,+)=db_{22}(\lambda,-)
       \nonumber \\ \kappa^{\ast}a_{12}(\lambda,+)=a_{21}(\lambda,-)
       \end{eqnarray}
       for all matrices $A_i,\; d_{\lambda}B_i$ in $\tilde{\Psi}_{sc}$
       \end{prop}

       The proof immediately follows from the fact that
       this permutation acts in the same way on all  our
       equations and substitutions including
       the original system in the variable $\lambda$
       and matrix $U$.

         Consider now the scalar function $\Phi$ which determines
         $\tilde{\Psi}_{sc}$ neglecting the terms of the order
          one in $\epsilon$
          \begin{eqnarray}
          \tilde{\Psi}^1_{sc}=\exp\{1/\epsilon B_{-1}+ B_0\}=
          R^{-1/4}b^{1/2}   \left(
          \begin{array}{cc}
           \Phi&0\\0&\Phi^{\ast}
           \end{array}\right)\nonumber \\
           d_{\lambda}\ln \Phi=\Omega=((1/\epsilon)\sqrt{R}
           +(a/b)_{\lambda}\frac{b}{2\sqrt{R}})d\lambda
           \end{eqnarray}

           From the definition of $\Omega,\; a,\;b,\;R$
           above we obtain immediately the following
          result:
           \begin{prop}
          The following 'Baker--Akhiezer' type asymptotic is true
          for the differential form $\Omega$ and function
          $\Phi$
          \begin{eqnarray}
          \Omega\sim \left(\frac{8}{\epsilon z^6}+\frac
          {C}{\epsilon z^2}-\frac {D}{4\epsilon}+O(z^2) \right) dz
          \nonumber \\
          z^{-2}=-\lambda=k^2 \nonumber  \\
          \Phi\sim \exp\left(-\frac{8}{5\epsilon} k^5-kx\right)
          \left(1-\frac {D}{4\epsilon}z+\frac{D^2}{32\epsilon}
           z^2
           +O(z^3)\right)
          \end{eqnarray}
          \end{prop}

          Consider now the analytical properties of the functions
         $\Phi$, $\Phi^2$ and \[\tilde{\Phi}=\sqrt{b} \Phi/\sqrt{-4}\]
           \begin{theorem}
           Let the equation ~\ref{eq:p11} be satisfied. In this case
           the following  properties are valid:
           function $\Phi$ is a well defined locally one-valued
            function on the double covering $\Gamma^{\ast}$ over the surface
            $\Gamma$ with the branching points  $R(\lambda)=0$
            and $((-u/2,
            \; \pm)$ determined by the function $R^{-1/4}b^{1/2}$;
           functions $\Phi^2$ and $\tilde{\Phi}$ are well-defined
          locally  one-valued meromorphic functions on $\Gamma$ and globally
          one--valued on some its covering
             with a free abelian
             monodromy group.
             The following formula is true
              for the function $\tilde{\Phi}$:
              \begin{eqnarray}
     \ln\tilde{\Phi}(w)=\frac{8}{5\epsilon w^5}-\frac{4g_2}{5
     \epsilon w}-\ln w+\frac{8g_3 w}{7\epsilon}-\frac{uw^2}{4}+
     O(w^3),\; w\rightarrow 0 \nonumber \\
     \tilde{\Phi}(w)=\frac{\sigma(w+a)}{\sigma (w) \sigma (a)}
     \exp\left \{-\frac{4}{5\epsilon} \wp\wp '(w)-\frac{4g_2}
     {\epsilon}\zeta(w)+\frac{6g_3w}{5\epsilon}-\zeta(a)w\right \}  \\
     \wp (w)=\mu=-\lambda,\; g_2=-\epsilon x,\; g_{3x}=
     \epsilon u/2 \nonumber
     \end{eqnarray}
     The equality
     \begin{eqnarray}
     d_w(\ln\tilde{\Phi})=\Omega+d_w(\ln b^{1/2}),\; \wp(w)=\mu \\
     -g_2=\epsilon x,\; g_{3x}=\epsilon u/2 \nonumber
      \end{eqnarray}
      is equivalent to the P--1 equation and gives
      nonlinear elliptic representation of P--1,
      depending on the parameter
       $w$.
     \end{theorem}

    Proof. The asymptotics of these functions near the
    infinite point follows from the proposition above.
    In particularly, the residue of the logarifmic
    derivatives of $\Phi^2$ in the variable $\lambda$
    is equal to zero. The corresponding logarifmic residue for
    the function $\tilde{\Phi}$ is equal to $-1$ by definition.
    Consider now the  singularities of these forms in the finite
    points of $\Gamma$. By definition we have
    \begin{eqnarray}
   d_{\lambda}(\ln \Phi^2)=2\Omega \nonumber \\
   d_{\lambda}(\ln \tilde{\Phi})=\Omega+\frac{b_{\lambda}}{2b}
   \end{eqnarray}

   Using the P--1 equation in the form
   \[u_x=2\sqrt{R}|_{\lambda=u/2}\]
   and the exact values of $a$ and $b$
   \[a=-u_x,\; b=2u+4\lambda\]
  we are coming to the following conclusion:  these
  differential forms may have poles in the points
  $(\mu,\pm)=(u/2,\pm)$ only, except infinity.

  It is obvious that $\Omega$ has the residues in both these points
  equal to $\pm 1/2$.

  The second form $\Omega+d_{\lambda}\ln b^{1/2}$ has a pole in
  one of these two points only, because the second term exactly
  cancels the pole on one sheet.  It has
  a nontrivial residue at infinity (equal to $-1$) coming from
  this second term.

All the residues of the both forms (who are the logarifmic
derivatives of the functions in the theorem )
are equal to $\pm 1$. Therefore the exponent
 from the integral for the both of these forms
 will be locally one--valued function
 on the Riemann surface $\Gamma$.

 Both of them will be obviously one--valued globally
 on some abelian covering over $\Gamma$.
 The theorem is proved.

 All these identities are valid for all solutions of the
 P-1 equation.
 \pagebreak

 For the {\bf Physical Solution} Novikov formulated some very
 special

 {\bf Conjecture}: There exists a function $F$ from the
 variables $g_2,\; g_3$  such that \[u=2\wp(F||g_2,\; g_3)\]

 Here we have \[g_2=-\epsilon x,\; g_{3x}=\epsilon u/2\]
 as above. For the real $x\rightarrow -\infty$ we should
 have     \[F(g_2,\; g_3)=(\omega)/2 \; +\delta\]

It is more convenient to write everything on the reduced
algebraic curve $\tilde{\Gamma}$ which has a good limit
   for $x\rightarrow -\infty$:\[g_2=t^4\tilde{g}_2=12,
   \; g_3=t^6\tilde{g}_3,\; F=t^{-1}\tilde{F}=\tilde{\omega}/2+
   \tilde{\delta}\]
   For the function $u$ we have: \[u=2t^2\wp(\tilde{\omega}+
   O((-x)^{-5/4})||12,\; \tilde{g}_3),\; t^2=(-\epsilon x)/12\]

   The limiting algebraic curve for $-x\rightarrow\infty$
   is rational \[y^2=4(s+2)(s-1)^2\]. For $\tilde{\Gamma}$
    we have \[y^2=4(s+2)(s-1)^2+O((-x)^{-5/4})\]
     for $\epsilon=1$.

   For $\tilde{g}_3$ we have a strictly negative
   correction which is coming from
   the first
    term $a^+_1$ in the asymptotic serie for the Physical
     Solution (see Introduction) \[\tilde{g}_3=-8+O((-x)^{-5/4}),
     \; \epsilon=1\]

 Therefore we have one real period $2\tilde{\omega}$
  of the lattice corresponding to
 the physical solution for $x\rightarrow -\infty$
  (it is generated by  two complex adjoint periods).

 {\bf Some good characterization of the Physical solution may come
    from the studying of the $x$--dependence of the
    muptiplicators of the function $\tilde{\Phi}$ along
     the global cycles on the Riemann surface
     $\tilde{\Gamma}$.}

\pagebreak
\vspace{7mm}

{\bf 4.Semiclassical approximation and foliations of the
Riemann surfaces.}

\vspace{5mm}

Consider a linear system \[\epsilon\Psi_t=V(t,\epsilon )\Psi\]
with small parameter $\epsilon$ and the right--hand
part of the form
\[V=V_0(t)+
V_1(t)\epsilon+\ldots\]
such that $V_0(t)$ is a polinomial or a rational function
in the variable $t$.

As it was pointed out above (see chapter 3), for the construction
 of the formal semiclassical solution we should at first
  reduce our system to the form which is diagonal
 in the first approximation.  After the substitution
 \[\Psi=U(t)\tilde{\Psi},\; U^{-1}V_0U=\tilde{V_0}=diag\]
we are coming to the admissible form
\[\epsilon\tilde{\Psi}_t=\tilde{V}\tilde{\Psi}=
(U^{-1}VU-\epsilon U^{-1}U_t)\tilde\Psi\]

The formal semiclassical solution was written in the form
\[\tilde{\Psi}_{sc}=(1+\sum_{i\geq 1}\epsilon^iA^i)
\exp\{\frac{B_{-1}}{\epsilon}+B_0+\sum\epsilon^iB_i\}\]

Here all  matrices $A_i$ have diagonal elements equal to zero,
 all matrices $B_i$ are diagonal. For simplicity
 we may think that all matrices here are of the order 2,
 and the trace of $V$ is equal to zero.

 The first diagonal term is
 \[\tilde{V}_0=diag(\sqrt{R},\; -\sqrt{R}),\; R=-\det V_0(t)\]

 Therefore this formula presents us some structure on the
  Riemann surface $\Gamma$ as above, the form
  \[\sqrt{R}dt=\Omega_0\]
  is meromorphic on the surface $y^2=R(t)$.

  Let us to ask:

  {\bf In which cases the formal semiclassical
  expression above  or at least its first two exponential terms
  \begin{eqnarray}
  \psi_{sc+}=\exp\left \{\frac{\int_{t_o}^t(
  +\sqrt{R(t')}
  +\epsilon b^{(1)}_{11})dt'}{\epsilon}\right \} \left\{ \left(
  \begin{array}{c}1\\0
  \end{array} \right)+O(\epsilon )\right\} \\
  \psi_{sc-}=\exp\left\{\frac{\int_{t_0}^t(-\sqrt{R(t')}+\epsilon
   b^{(1)}_{22})dt'}{\epsilon}\right\} \left\{ \left(
  \begin{array}{c}0\\1
  \end{array}\right)+O(\epsilon )\right\}
  \end{eqnarray}
  gives the right asymptotic in $\epsilon$ for some exact solution
  $\psi_{+}(t,\epsilon)$ along all the path of integration?}

  The following theorem gives an answer for this question.
   \begin{theorem}
   Consider a hamiltonian foliation of the surface $\Gamma$
   given by the zeroes of the harmonic form  which is a real part of
   the abelian differential 1--form on the Riemann surface $\Gamma$
   \begin{equation}
   \omega=0,\; \sqrt{R}dt=\omega+i\omega '
   \label{eq:fol}
   \end{equation}
   and two different orientations in it:
   The orientation $+$ is such that \[\int_{t_0}^t\omega\]  increases
    along this path in its positive direction; the opposite
    orientation is $-$.
    The formulas above  describe the right asymptotic for some
    exact solutions
    increasing along the positive path of integration $\gamma$
    and increasing for $\epsilon\rightarrow 0$ if this path is
    transversal to foliation in all its points. It should be
    oriented in the direction of the increasing of
    the function $\int_{t_0}^t\omega$ for the $\psi_{sc+}$
    and in the opposite direction for $\psi_{sc-}$.
    Two different paths with
    the same endpoints describe the same solution along the union
    of these paths if they are homotopic in the class of
    paths everythere transversal   to foliation with fixed endpoints.
    \end{theorem}

    {\bf Remark 1. By definition the decreasing asymptotics
    are the same one but directed in the opposite
    way.}

    {\bf Remark 2. If there is no small parameter $\epsilon$
     in the linear system above, we may use a variable $t^{-1}$
     as a small parameter in many cases and  apply this statement
     for the large $t\rightarrow\infty$}.

   There is no need to give a proof for this statement. In fact,
   it is known many years (in different terminology) for special cases
   in the physical and mathematical literature (see \cite{fed}),
   but things never had been put in this terminology, connecting this
   problem with the theory of  algebraic
   Riemann surfaces and the topology of these beautiful
   and nontrivial foliations.

   {\bf I believe, this form gives
   a beautiful motivation for the studying very nontrivial topology
   of special Hamiltonian
   foliations on the Riemann surfaces given by the real parts
   of holomorphic and meromorphic forms.}

  Even for the simplest generalization of the straight --line flow
   on the torus as a foliation generated by the real part
    of holomorphic 1-form on hyperelliptic Riemann surface of genus 2
    nothing has been done.

    \pagebreak

  {\bf Example 1.} Let $R(t)$ be a polinomial of the odd degree
  $n=2m+1$.

  We have a following foliation near the infinite point (see
  Fig 1).  There exist exactly $2(2m+3)$ sectors near the point
  $v=\infty,\; v^2=t{-1}$  on the Riemann surface near infinity,
  which exactly cover the $2m+3$ sectors  on the $t$--plane,
  separated by 'separatrices':
  \[t^{-1}=\exp\{\rho+i\phi\},\; 2k\pi/j\leq \phi\leq 2(k+1)\pi/j,\;
   k=0,1,2,\ldots,2m+2\]

   \pagebreak

  In each sector we have a
  picture  like in Fig 1. Each trajectory starts in
  the infinite point $t^{-1}=0$ asymptotically
  tangent to the separatrix and turns
  to anover
  separatrix on the boundary of this sector. Finally it returns
  to the infinite point asymptotically tangent to the last separatrix.
  Let us introduce the orientation (as above) in this foliation.

  We may observe that the neighboring sectors have the opposite
   orientations: if some transversal path is directed to the $\infty$
    in some sector  and it is positive, so in the neighboring sectors
     the positive transversal paths will be directed out of the
     $\infty$.

  Starting from some point
  $t_0$ ('very closed to infinity')
  in the sector (i) we may come by the transversal
  paths  $\gamma_{i+1}'$ or by  $\gamma_{i-1}''$ for the proper
  orientation
  $\pm$  (which admits the positive transversal
  path out of $\infty$ in this
   sector)
 to the neighboring sectors $(i-1)$ and $(i+1)$
  and to follow along these paths in the direction
  of the infinite point. Therefore the solution which increases
   along these paths (and decreases in the inverse direction
   to the infinite point in the sector $(i)$) has a natural continuation
   in the neighboring sectors. We shall denote it by $\psi_i$
\begin{prop}
   There is a 'Stokes relation' between the solutions:
   \begin{equation}
   \psi_{i+1}-\psi_{i-1}=\alpha_i\psi_i
   \label{eq:stok}
   \end{equation}
   Here $\alpha_i=$const such that \[\alpha_i=\alpha_{i+2m+3}\]
   \end{prop}

   The proof of this relation follows from the fact that there
   is exactly one--dimensional family of solutions in each sector which
    have a decreasing behavior for $t\rightarrow\infty$. Both
     solutions $\psi_{i\pm 1}$ have a continuation in the sector
     $(i)$ as the increasing ones, described by the same
     asymptotics. Their difference is proportional to the decreasing
     solution in this sector with some  constant coefficient because
     the system has an order  two. Therefore the first part is proved.

     By the origin of this
     system on $\Gamma$ from the original $t$-plane we have
     the same(or proportional) solutions over the sectors
     which cover the same domain
      in the $t$-plane up to permutation of the matrix elements.
      From this the proof of the last relation between
      the Stokes coefficients follows immediately.

     \pagebreak

     {\bf Example 2.} Consider now the asymptotics for the Physical and
     Nonphysical solutions for $x\rightarrow-\infty$ as above:
     \[u_{\pm}\sim\pm\sqrt{\frac{-\epsilon x}{3}}\]

     After the substitution
     \begin{eqnarray}
     x=-\tau^{4/5},\;u=\frac{\tau^{2/5}}{\sqrt{3}}\tilde{u}(\tau)
     \nonumber \\
     \lambda=-\frac{\tau^{2/5}}{2\sqrt{3}}\mu \nonumber \\
     \tilde{\psi}_1=\psi_1,\; \psi_2=-\frac{5}{4}\tau^{1/5}
     \tilde{\psi}_2
     \end{eqnarray}
     we are coming to the system
     \[\frac{1}{\tau}\tilde{\Psi}_{\mu}=\tilde{\Lambda}_{\pm}\tilde{\Psi}\]
     such that
     \begin{eqnarray}
     \tilde{u}=\pm 1+o(1)\nonumber \\
     \tilde{\Lambda}_{\pm}=-\frac{1}{12}\left(\begin{array}{cc}
     0&-5(\pm 1-\mu) \\
     -\frac{8}{5\sqrt{3}}(3-\mu^2-\mu(\pm 1)-1)&0
     \end{array}\right)+O\left(\frac{1}{\tau}\right)
     \end{eqnarray}

     Here the quantity $\tau^{-1}$ plays a role of small parameter.
      Riemann surface has a genus, equal to $0$.
     The corresponding foliations you may see on the Fig 2
     (for the Physical
     Solution)
     and on the Fig 3 (for the Nonphysical one).

     \pagebreak

     We have exactly two finite critical points of these foliations in the
     following points (only one of them is nondegenerate which is
      located in the points
     $\mu=\pm 1$):
     \[\mu=1,\; \mu=-2\]

     for the Physical solution (there is no separatrices
      joining two finite critical points here. All trajectories
       including separatrices
      are coming to infinity).
     \[\mu=-1,\; \mu=2\]
     for the Nonphysical Solution (there exist a separatrix line joining
     two finite critical  points  along the real line.
     All other trajectories including separatrices
      are coming
     to infinity.)

     \pagebreak
\vspace{7mm}

{\bf 5.Special semiclassics for the Lax pairs in the case of Physical
 Solution}
 \vspace{5mm}

   We shall start our studying of the Physical solution from the
   substitution (rescaling) described in the formulas (31)
   at the end if the
   paragraph 4 using the same notations \[\tilde{u},\;
    \mu,\;\tau,
   \tilde{\Psi}=(\tilde{\psi}_1,\tilde{\psi}_2),\; \tilde{\Lambda},\;
   \tilde{Q}\]

   Let us write more explicit formulas
   \begin{eqnarray}
   \tilde{\Lambda}=-\frac{1}{12}\left(\begin{array}{cc}
   \frac{2\tilde{u}(\tau) +5\tau\tilde{u}'(\tau)}{2\tau}&
   5(\mu-\tilde{u}(\tau)) \\
   -\frac{8(3-\mu^2-\mu\tilde{u}(\tau)-\tilde{u}^2)}{5\sqrt{3}}&
   \frac{-2\tilde{u}(\tau)-5\tau\tilde{u}'(\tau)}{2\tau}
   \end{array}\right)=\nonumber \\
   =-\frac{1}{12}\left(\begin{array}{cc}
   0&5(\mu-1) \\
   -\frac{8(3-\mu^2-\mu-1)}{5\sqrt{3}}&0
   \end{array}\right)+O\left(\frac{1}{\tau}\right) \nonumber \\
   \tilde{Q}=\left(\begin{array}{cc}
   0&1\\
   \frac{16\tilde{u}+8\mu}{25\sqrt{3}}&-\frac{1}{5\tau}
   \end{array}\right)+\frac{2}{5}\mu\tilde{\Lambda}\nonumber \\
   \frac{1}{\tau}\tilde{\Psi}_{\mu}=\tilde{\Lambda}\tilde{\Psi},\;;
   \tilde{\Psi}_{\tau}=\tilde{Q}\tilde{\Psi}
   \end{eqnarray}

  This is a very interesting special Lax Pair (or Zero--Curvature
  Representation) such that its second equation is written
  in terms of the parameter $\tau$ or $\tau^{-1}$ which
plays a role of the small parameter ('Plank constant') for
the first equation of this pair written in the variable $\mu$.

{\bf Especially interesting point here which is valid for the Physical
 Solution only is the following: there exist an asymptotic serie (3) for
 this  solution $\tilde{u}_+$ in the variable $\tau$.
 Therefore we shall try to
  look for the Semiclassical Solution $\tilde{\Psi}_{sc}(\mu,\tau)$
  for the $\mu$--equation of
   this pair which is a formal serie in the variable $\tau^{-1}$
   and satisfies
    to the $\tau$--equation also.}

 More precisely, after the substitution below we are coming as before
 to the equations which are diagonal modulo $\frac{1}{\tau}$:
 \begin{eqnarray}
 \tilde{\Psi}=U\Phi,\; U=\left(
 \begin{array}{cc}
 1&1 \\
 -\sqrt{\alpha(\mu+2)}&\sqrt{\alpha(\mu+2)}
 \end{array}
 \right) \nonumber \\
 \alpha=\frac{8}{25\sqrt{3}},\;
 \frac{1}{\tau}\Phi_{\mu}=\bar{\Lambda}\Phi,\;
 \Phi_{\tau}=\bar{Q}\Phi \nonumber \\
\bar{\Lambda}=\left(\begin{array}{cc}
r_+(\mu)&0 \\
0&r_-(\mu)
\end{array}\right)+O\left(\frac{1}{\tau}\right) \nonumber \\
\bar{Q}=\left(\begin{array}{cc}
q_+(\mu)&0 \\
0&q_-(\mu)
\end{array}\right)+O\left(\frac{1}{\tau}\right)
\end{eqnarray}

It is useful to observe that \[q_+(\mu)=\int_{-2}^{\mu}r_+(\mu')
d\mu'=\int_3^{\mu}r_+(\mu')d\mu'\]

Here we have
\begin{equation}
q_{\pm}(\mu)=\frac{\pm\sqrt{\alpha}}{6}(\mu-3)(\mu+2)^{3/2},\;
r_{\pm}(\mu)=\pm\frac{5}{12}(\mu-1)\sqrt{\alpha(\mu+2)}
\label{eq:alg}
\end{equation}

Let us formulate the following main theorems of this section:
\begin{theorem}
The last 'rescaled' Lax pair  (34) admit a unique formal quasiclassical
 matrix--solution of the form
 \begin{equation}
 \Phi_{sc}=\tau^{-\frac{1}{10}}(\mu+2)^{-1/4}
 (1+\sum_{i\geq 1}^{\infty}\frac{A_i}{\tau^i})
 \exp\{\tau B_{-1}+\sum_{i\geq 1}^{\infty}B_i\tau^{-i}\}
 \label{eq:scalg}
 \end{equation}
 such that all $A_i$ are off--diagonal, all $B_i$ are diagonal,
  $A_i$ and $B_i$ are algebraic functions on the Riemann surface
  $\Gamma^0$
  of genus zero
  \begin{equation}
  \Gamma^0=\{y^2=\mu+2\}
  \label{eq:rim}
  \end{equation}
  For the most important term we have
  \begin{equation}
  B_{-1}=\left(\begin{array}{cc}
  q_+(\mu)&0 \\
  0&q_-(\mu)
  \end{array}\right),\; q_-=-q_+ ,\;\; B_0=0
  \label{eq:sc1}
  \end{equation}
  \end{theorem}
  \begin{theorem}
  Consider the {\bf 'Exact Stokes Sectors'} bounded by the lines
  \begin{equation}
  Re[q_{\pm}(\mu)]=0
  \label{eq:est}
  \end{equation}
  on the Riemann surface $\Gamma^0$ (see Fig 4).
  For each Exact Stokes Sector there
   exist a unique up to a constant factor exact vector--solution
   for the rescaled
   Lax pair (34) which is decaying as $\tau\rightarrow\infty$.
   Its asymptotics may be described by the first exponential
   term of the semiclassical
   solution (modulo terms of the order $O\left(\frac{1}{\tau}\right)$
   inside and outside of the exponent)
   in the previous theorem.
   \end{theorem}

\pagebreak

We may see on the Fig 4 that there are two Exact Stokes Curves (39)
 on the Riemannian surface $\Gamma^0$
 which are closed and project  on the closed curves in  the
  extended $\mu$--plane plus infinity. They are
  passing through the points $(\mu=3,\pm)$ and $\infty$. Asymptotically
   near infinity these two closed curves give us  four
   different asymptotic Stokes lines with the numbers 1 and 2
  (which bound a sector number 2) and with the numbers 6 and 7
  (which bound a sector number 7). A Stokes Sector number $k$
   is bounded by the lines number $(k-1,\; k),\; k=1,2,\ldots,10.$

  Other six  Exact Stokes Lines project in  the graph on the
  $\mu$--plane which contains 3 curves starting from the point $\mu=-2$
   in the directions of the angles $\pi,\; \pm \frac{\pi}{3}$.
   One of them is exactly a negative part of real line
   below the point $\mu=-2$.
   The total preimage of this graph on $\Gamma^0$ gives
   the Exact Stokes Curves which
   asymptotically near infinity exactly coinside with the Stokes
    lines number 3 and 8, 4 and 9, 5 and 10
    (see numeration on the Fig 4).

   \pagebreak

  From the theorem 5 and this picture we have immediately the following
  conclusions:
  \newtheorem{cor}{Corollary}
  \begin{cor}
  The Stokes coefficients defined in the Proposition 6
  are constant (they don't depend on $\mu,\;\tau$);
  the Stokes coefficients number 2 and 7 are equal to zero
  \[\alpha_2=\alpha_7=0\]
  For other coefficients we have \[\alpha_k=\alpha_l,\;k-l=0(mod 5)\]
  \[\alpha_j=i,\; j=4,\; 5,\; 9,\; 10\]
  \[\alpha_1+\alpha_3=\alpha_6+\alpha_8=i\]
  \end{cor}

  The proof of this corollary is the following. The ordinary
  Stokes Sectors with the
  numbers 1 and 3 in fact are connected with each other because they
  belong to the same domain (39). Therefore
  we have $\phi_1=\phi_3$ for the appropriate solutions which have
  a $\tau$--decay in the sectors 1 and 3 correspondingly and $\alpha_2=0$
  by definition. We have also $\alpha_2=\alpha_7$--see proposition 6.

  Let us eliminate now the closed exact Stokes lines from the
  Riemann surface $\Gamma_0$. Consider the following functions
  $F_k$:
  \[\phi_1=\phi_3=F_1,\; \phi_4=F_2,\; \phi_5=F_3,\; \phi_6=
  \phi_8=F_4,\; \phi_9=F_5,\; \phi_{10}=F_6\]

For the last solutions $F_k$ we have practically the same
Stokes coefficients
but numeration is modulo 6.
\[F_{k+1}-F_{k-1}=\beta_kF_k\]
\[\beta_k=\beta_l,\;F_k\sim F_l,\; k-l=0(mod 3)\]
\[\beta_k=i=\sqrt{-1}, \;  k=1,2,3,4,5,6\]
They coinside with the Stokes coefficients
for the equation describing the Airy function where the infinite
point meets exactly six Stokes sectors. This is true
just because it follows from the algebraic symmetry:
\[\beta_{k+3}=\beta_k,\; F_{k+3}\sim F_k\]
and from the property that semiclassical solution has an $\infty$
 as a double branching point (see a theorem 2).
 It means that we have to put sign
 $-1$ after the passing the full angle $2\pi$ around $\infty$.

 For the proof we are writing
 a sequence of formulas: \[F_2=F_0+\beta_1F_1,\;
 \phi_3=F_1+\beta_2F_2,\ldots\]
 and using this algebraic symmetry. The full angle we shall pass
 after the six steps. Expressing the functions $F_k$
  through the previous ones and finally through the $F_1,\; F_2$
   we shall get the algebraic equations for the coeffitients $\beta_k$.
   Our formulas for them  follows automatically.

By definition we have: \[\alpha_k=\beta_{k-2}=\beta_{k+1},\; k=4,5\]

We have also \[\beta_1=\alpha_1+\alpha_3\]
because \[F_2-F_6=\phi_4-\phi_{10}=(\phi_4-\phi_2)+(\phi_2-\phi_{10})=
(\alpha_1+\alpha_3)\phi_1\]
The corollary is proved.

{\bf Remark.} In the work \cite{kop} (see also \cite{kk,kit}) this result
about the Stokes coefficients for this special solution has been
found. These authors used different (more complicared) Lax--type
representation for the P--1 equation extracted from \cite{m-j},
but the most important difference from our work is the following:

  the authors of \cite{kop,kk,kit} and others in St--Petersburg group
 did not worked with the common solution for the both equations
 for  the Lax pair. They worked with one linear system (which is
 in the variables $\lambda$ or  $\mu$ after rescaling) and used the
 following fact:  Stokes coefficients don't depend on $\tau$ or $x$ iff
 $u$ satisfies to the P--1 equation. It means in fact that these authors
 worked with picture like in Fig 2 and had a need to investigate
 the neighborhood of the critical point $\mu=1$ of our foliation
  and to prove that it has in fact a trivial monodromy.

  {\bf In our approach this difficulty does not exist because this point is
   nonsingular at all for the special semiclassics for the
   (rescaled) Lax pair in the case of the Physical Solution.}

Let us consider now the behavior of the semiclassical solution in the
interval of real line between the points $\mu=-2$ and $\mu=3$.
 It is the most important interval because its small  shift
 in the variable $\mu$ in the negative imaginary direction connects two
  Exact Stokes Sectors: the sector number 2 bounded by the separatrices
   number 1 and 2 near infinity (or by 6 and 7 on the second sheet)
    and the sector number 0 bounded by the separatrices number 9 and 10
    near infinity (or by 4 and 5 on the second sheet)---see Fig 4.

There is only one essentially nontrivial Stokes coefficient
 $\alpha_1$ which determines the difference between
 two Physical Solutions in the one-parametric family of them.
It is exactly equal to the difference between the solutions
$\phi_0$ and $\phi_2$ divided by the solution $\phi_1$.
These special solutions $\phi_i$ have decay in the corresponding
 exact Stokes sectors in the variable $\tau$ and decay in the
 variable $\mu$ if $\mu\rightarrow\infty$ in this sector.

 {\bf The points $\mu=-2$ and $\mu=3$ are singular for the $\tau$--
 dynamics in the pair of the Systems (33)}.

 We are starting from the original
 nondiagonal form (33) because the transformation (34)
  is singular in the point $\mu=-2$. Modulo $\tau^{-2}$
  we have:
 \begin{eqnarray}
 \tilde{\Psi}_{\tau}\tilde{\Psi}^{-1}=\frac{-1}{6}
 (\mu-3)(\mu+2)\left(\begin{array}{cc}
0&1 \\ \alpha(\mu+2)&0
\end{array}\right)+ \nonumber \\
+\left(\begin{array}{cc}
-\frac{\mu}{30\tau}&0 \\
0&\frac{\mu-6}{30\tau}
\end{array}\right)+O(\frac{1}{\tau^2})
\end{eqnarray}

After the substitutions \[\mu=3,\;  \mu=-2\]
we are coming to the following proposition
\begin{prop}
The exact solution $\tilde{\Psi}$ in the variable $\tau$
has a behavior like some power of $\tau$ for $\tau\rightarrow\infty$
in the points $\mu=-2$ and $\mu=3$. Consider the special solution
\[\tilde{\psi}_1=U(\mu)\phi_1 \]
for the system (33)
which has an exponential decay for $\tau\rightarrow\infty$
in the Exact Stokes Sector
containing the open interval $(-2,\;  3)$ in the variable $\mu$
 by the theorem 5. Here $U$ is given by the formula (34).
 This solution is increasing for
$\mu\rightarrow -2$ and $\mu\rightarrow 3$ for $\tau$
 large enough.
 \end{prop}

 Proof of this statement is obvious.

The following result might be immediately extracted from this:
\begin{cor}
Consider the Linear Operator \[\hat{A}=\frac{1}{\tau}\partial_{\mu}
-\tilde{\Lambda}(\mu,\; \tau)\]
The spectral problem \[\hat{A}\psi=\lambda\psi\] with the 'Plank parameter'
$h=\frac{1}{\tau}$
has a point $\lambda=0$ as a discreet semiclassical
eigenvalue for the Dirichle
Problem on the interval
$(-2,\; 3)$ in the variable $\mu$. It means that the boundary conditions
are satisfied modulo  exponentially small  nonaccuracy in the endpoints
$\mu=-2,\; 3$.
\end{cor}

Proof. Consider any solution $\tilde{\psi}$
for the rescaled Lax Pair (33) in the variables
 $\mu,\; \tau$ different from the decreasing
 solution $\tilde{\psi}_1$ above.
This solution
  has an increasing asymptotics
  for $\tau\rightarrow+\infty$ because it has a
  semiclassical behavior written in the theorem 4
  which obviously describes its asymptotics (at least its first main
  exponentially increasing term does). Therefore the modulus of this solution
   has a maximum somewhere
   near the point \[\max_{-2\leq\mu\leq3}q_+(\mu)\;\; =1\]
   Its growth will be less and less for $\mu\rightarrow -2$
   and $\mu\rightarrow 3$. After the proper normalization  of $\psi$
   (making the maximum of modulus  equal to 1 or the norm in $L^2[-2,\;\;3]$
    equal to 1) we shall see that the normalized solution of the
   equation $\hat{A}\psi=0$ has some decay for $\mu\rightarrow -2,\;3$
   and therefore may be treated as a semiclassical eigenfunction
   corresponding to the Dirichlet Problem on this interval.
   The corollary is proved.

   {\bf Conjecture. The property of the Operator $\hat{A}$ described
   in the  previous Corollary determines all formal serie (3) in the Physical
    Solution.}

\pagebreak

   \vspace{7mm}
   {\bf 6.Proof of the main theorems of the p.5}

  \vspace{5mm}

   We shall give in this paragraph the complete proof for
   the theorems 4 and 5.

   {\bf Proof of the theorem 4.}

   Let us  write the system (34)
   for the matrix $\Phi$ in the following form:
   \begin{eqnarray}
   \Phi_{\mu}=\tau (V_0+V_1^d\tau^{-1}+V_1^a\tau^{-1}+V^{rest})\Phi\nonumber \\
   \Phi_{\tau}=(W_0+W_1^d\tau^{-1}+W_1^a\tau^{-1}+W_{rest})\Phi\nonumber \\
   W_{rest}=\sum_{i\geq 2}W_i\tau^{-i},\; W_i=W_i(\mu)
   \end{eqnarray}
   Here 'd' means diagonal and 'a' means offdiagonal parts of the
   matrices,  $W_0$ and $V_0$ are already diagonal.

   Our strategy now is to solve this system in the variable $\tau$
   instead of  $\mu$ as a formal serie.
   it is easy to check the following formulas for the solution
    (36):
   \begin{eqnarray}
   B_{-1}=W_0,\; A_1B_{-1}=W^a_1+W_0A_1,\; B_0=0,\; \nonumber \\
   A_2B_{-1}=W_0A_2+A_1+B_1+W_1^aA_1+W_2,\;\; \ldots \nonumber \\
   A_nB_{-1}-B_{-1}A_n=(n-1)B_{n-1}+ \nonumber \\
   +P_n(W_1^a,W_2,\ldots ,W_n,
   A_1,\ldots ,A_{n-1},B_1,\ldots ,B_{n-2})
   \end{eqnarray}
   Here $P_n$ are some noncommutative polinomials.

   From this formulas we deduce that the solution in the form of the
    desired formal serie in the variable $\tau$  exists and is
    unique. In particularly we have:
    \begin{eqnarray}
    A_1=Ad[W_0]^{-1}(W_1^a),\; Ad[X](Y)=[X,Y]\nonumber \\
    B_n=-n^{-1}P_{n+1}^d,\; A_n=Ad[W_0]^{-1}(P^a_n)
    \end{eqnarray}

   Let us now substitute this formal serie in the first equation
    of the system (34) in the variable $\mu$. Consider
    the expression \[X(\mu,\tau)=(\partial_{\mu}-\tau\bar{\Lambda})
    \Phi_{sc} \]
    Obviously we have:
    \[\tau^{\frac{1}{10}}\mu^{1/4}X=
    (c_1\tau^{-1}+c_2\tau^{-2}+\ldots)\exp\{B_{-1}\tau+B_1\tau^{-1}+
    \ldots\}\]
    Let us remark now that $X$ also satisfies to the equation
     in $\tau$ of the  system (34), as $\Phi_{sc}$, because
     the operators $\partial_{\mu}-\tau\bar{\Lambda}$
      and $\partial_{\tau}-\bar{Q}$ commute with each other.

    Adding $X$ to the solution $\Phi_{sc}$ above we shall get a new
    solution \[\Phi_{sc}+X=\tau^{\frac{-1}{10}}\mu^{-1/4}
    (1+\frac{A_1+c_1}{\tau}+\ldots)
    \exp\{B_{-1}\tau+B_1\tau^{-1}+\ldots\}\]
    which is easily to rewrite in the admissible form
    \[\Phi_{sc}+X=\tau^{\frac{-1}{10}}\mu^{-1/4}
    (1+\frac{A_1+c_1^a}{\tau}+\ldots)\exp\{B_{-1}\tau+
    (B_1+c_1^d)\tau^{-1}+\ldots\}\]
    like the solution $\Phi_{sc}$.
    From the uniqueness of the admissible semiclassical
    solutions in this form
     we deduce that \[X=0\]

     {\bf Therefore the theorem 4 is proved.}

   \pagebreak

{\bf Proof of the theorem 5.}

Consider now any exact Stokes Sector in which we have
\[Re[q_+(\mu)]\geq 0\]
Let us introduce for any point $(\mu,\pm)$ in this sector
a special semiclassical solution for the equation (42)
in the variable $\tau$:
\begin{equation}
\phi_{dec}=\tau^{-\frac{1}{10}}\mu^{-1/4}\left[\left(
\begin{array}{c}
0\\1
\end{array}
\right)+\sum_{i\geq 1}\phi^-_i\tau^{-i}+\ldots\right]e^{q_-(\mu)\tau}
\label{eq:scdec}
\end{equation}

Let \[\tilde{\phi}=\tau^{-\frac{1}{10}}\phi\]
and
\begin{eqnarray}
L_0\tilde{\phi}=\left[\frac{W_1^a}{\tau}+W_{rest}\right]\tilde{\phi}
\nonumber \\
L_0=\partial_{\tau}-W_0
\end{eqnarray}

The corresponding Green function of the Operator $L_0$
may be written in the form
\begin{eqnarray}
G_{dec}=\left\{\begin{array}{c}
\left(\begin{array}{cc}
-1&0\\
0&0 \end{array}\right)e^{(\tau-y)q_+(\mu)},\; \tau <y \\
\left(\begin{array}{cc}
0&0\\ 0&1 \end{array}\right)e^{(\tau-y)q_-(\mu)},\; \tau >y
\end{array}\right. \\
L_0G_{dec}=\left(\begin{array}{cc}
1&0\\ 0&1 \end{array}\right)\delta(x-y)
\end{eqnarray}

\newtheorem{lemma}{Lemma}
\begin{lemma}
For any point $\mu$ inside of the Exact Stokes Sector
$Re(q_+(\mu)>0$ and $\tau_0$ sufficiently large
there exists  a unique exact solution $\phi_{dec}(\tau,\;\mu)$
for the equations (34),
(41) in the variable $\tau$
such that it has an asymptotics (44) above for $\tau\rightarrow+\infty$.
\end{lemma}

{\bf Proof of Lemma 1.} We write the equation for finding  the
 desired solution in the form of Integral Equation:
 \begin{eqnarray}
 \tilde{\phi}=\left(\begin{array}{c}
 0\\1 \end{array}\right)e^{\tau q_-(\mu)}+T_{dec}\tilde{\phi}
 \nonumber \\
 (T_{dec}f)(\tau)=\int_{\tau_0}^{\infty}G_{dec}(\mu,\tau,y)\left[
 \frac{W_1^a}{y}+W_{rest}(\mu,y)\right]f(y)dy
 \end{eqnarray}

 Here $f$ is a vector--function \[f=\left(\begin{array}{c}
 f_1\\f_2\end{array}\right)\]

 Let us introduce a norm \[||f||_c=\delta+c\kappa\]
 depending on the parameter $c>0$. Here we have by definition:
 \[\delta=\sup_{\tau_0\leq\tau}|f_2(\tau)e^{\tau q_+(\mu)}|<\infty\]
 \[\kappa=\sup_{\tau_0\leq\tau}|\tau f_1(\tau)e^{\tau q_+(\mu)}|<\infty\]

 Therefore we have \[|f_1(\tau)\leq \frac{c_1}{c\tau}|e^{-\tau q_+(\mu)}|\]
 \[|f_2(\tau)|\leq c_1|e^{-\tau q_+(\mu)}|\]
 if $||f||_c=c_1$.

 Inside of the Exact Stokes Sector we may estimate the coefficients
 of the equations (34), (42). For $\tau_0$ large enough we have:
 \[\left|\frac{W^a_1}{\tau}+W_{rest}\right| \leq \left|
 \begin{array}{cc}
 \frac{d_1}{\tau^2}&\frac{d_2}{\tau}\\
 \frac{d_3}{\tau}&\frac{d_4}{\tau^2}\end{array}\right|\]
 where $d_i$ are some negative constants (depending on $\mu$ in principle).

  This follows in the
 obvious elementary way from
  the formulas for these matrices.

 Applying the operator $T_{dec}$ to the vector $f$
  with norm $||f||_c=c_1$ we are coming to the following
 inequalities:
 \begin{eqnarray}|g_1|\leq c_1\int_{\tau}^{\infty}
 \left|e^{q_+(\mu)(\tau -2y)}\right|dy
 \left(\frac{d_1}{c\tau^3}+
 \frac{d_2}{\tau}  \right)=\nonumber\\
 =c_1\left|e^{-\tau q_+(\mu)}\right|
 \frac{1}{2Req_+(\mu)}\frac{1}{\tau}\left(d_2+
 \frac{d_1}{c\tau^2}\right)\nonumber \\
 |g_2|\leq c_1\int_{\tau_0}^{\tau}\left(\frac{d_3}{c}+d_4\right)
 \frac{dy}{y^2}\left|e^{\tau q_-(\mu)}\right|\leq \nonumber\\
 \leq \frac{c_1}{\tau_0}\left|e^{\tau q_-(\mu)}\right|
 \left(\frac{d_3}{c}+d_4\right)
 \end{eqnarray}

For the norm of the image $g=T_{dec}f$ we obtain from (49) inequality:
\begin{equation}
\left|\left|\begin{array}{c}g_1\\g_2\end{array}\right|\right|_c
\leq \left(\frac{c}{2Req_+(\mu)}\left(d_2+\frac{d_1}{c\tau_0^2}\right)+
\frac{1}{\tau_0}\left(\frac{d_3}{c}+d_4\right)\right)\left|\left|
\begin{array}{c}
f_1\\f_2\end{array}\right|\right|_c
\end{equation}

Choosing a constant $c$ small enough  and $\tau_0$ big enough
we see that the norm of Operator $T_{dec}$ is small. Therefore
the Operator $T_{dec}-1$ is invertible and the Integral equation (48)
above is uniquely solvable.

{\bf Lemma 1 is proved.}

We know now that in any point $\mu$ inside of the Exact Stokes Sector
there exists a special one-dimensional subspace of the solutions
proportional to $\phi_{dec}$
in all 2--dimensional
linear space of solutions of the equation (41) in the variable $\tau$
constructed in the lemma 1 above.
\begin{lemma}
Consider any solution $\phi(\mu,\tau)$ for the
system (34), (41) in the variable
$\mu$ inside of some Exact Stokes Sector such that it coinsides
 with the special solution $\phi_{dec}$ in one point $\mu_0$.
In this case the solution $\phi$ is proportional to  $\phi_{dec}$
everythere inside of the Exact Stokes Sector.
\end{lemma}

{\bf Proof of Lemma 2.} Let us compare two points $\mu_0,\;\mu_1$
inside of this sector which are connected by the 'negative'
 path $\mu(t),\; 0\leq t\;\leq 1$,
  inside this sector with the property
  \[Re\left[r_+(\mu(t))\frac{d\mu}{dt}\right]\leq 0\]

  Starting from the solution
 $\phi_{dec}(\mu_0,\tau)$ we find the solution $\phi(\mu,\tau)$
 for all the points on this curve and for all $\tau\geq \tau_0$
 solving the equation (34),(41) along the variable $\mu$ with
 fixed $\tau$ large enough.
  It is possible because
   the equations in $\mu$ and $\tau$ commute with each other.

   We have an obvious estimate for the equation along the curve:
   \[||z(t)||\leq ||z(0)||\; \;||A(t)||\]
   where
   \[z(t)=\phi(\mu(t),\tau)\]
   and $A(t)$ is the operator of evolution generated by the linear
    system (34), (41) along the curve  $\mu(t)$
   \[\frac{dz}{dt}=\left(\tau\bar{\Lambda}\frac{d\mu(t)}{dt}\right)z(t)\]
   after the diagonalization of the first nontrivial term.

   There exists such  constant $c$ that the following
   inequality is true for the  operator of evolution
    along the 'negative' path:
   \[||A(t)||\leq \exp\{\int_0^t \left(\tau \left | Re\left[r_+(\mu)
   \frac{d\mu}{dt} \right] \right | +c\right)dt\}\]
   because our matrix in the left side of the $\mu$--equation
   along the curve
   is diagonal modulo terms of the order 0.
    We know that the initial   value is exactly the
    solution $\phi_{dec}(\mu_0,\tau)$.

      Starting from the point $\mu_0,\tau$
     and from $\phi_{dec}=z(0)$ as an initial value we apply the
     inequality above for the norm $||z(t)||$. We are coming to the
    following estimate
    \[|\phi(\mu,\tau)|\leq c_2\exp\left\{-Re[q_+(\mu_0)]\tau+ \left |
    Re\left[ \int_{\mu_0}^{\mu}r_+(\mu')d\mu'\right] \right | \tau\right\}
    \leq \] \[\leq c_2\exp(-\tau Re[q_+(\mu)])\]
     for the solution under investigation and for any $\tau$ large enough.

    Combining this with the previous inequality we are coming
    to the following conclusion:
     this solution exactly  coinsides with the solution $\phi_{dec}(\mu,
     \tau)$
    in the variable $\tau$ because it has a right asymptotics
    for $\tau\rightarrow\infty$ in the point $\mu$.
    (Let us to point out that any other solution $\phi(\mu,\tau)$
    in fact has an exponential growth in $\tau$ for
    $\tau\rightarrow\infty$ inside of this Exact Stokes Sector.)

    Consider now an arbitrary point   $\mu_1$ in the same sector.
    We know already that for all such points $\mu_1$ which
    may be connected with $\mu_0$ by the 'negative' path $\mu(t)$
    our lemma is true. This lemma therefore is true for 'positive'
    paths also changing the role of $\mu_1$ and $\mu_0$ and
    using the uniqueness of the solution with decay in $\tau$ for
    any given $\mu$. We may connect any 2 points in the sector
    by the finite product of 'positive' and 'negative' paths.
    This gives the proof of lemma for our sector.

     Exact Stokes Sectors with $Re[q_-(\mu)]>0$  may be
    considered in the same way up to inessential changes.

    {\bf Lemma 2 is proved.}

    Proof of the theorem 5 follows from the lemmas 4 and 5.

    \pagebreak

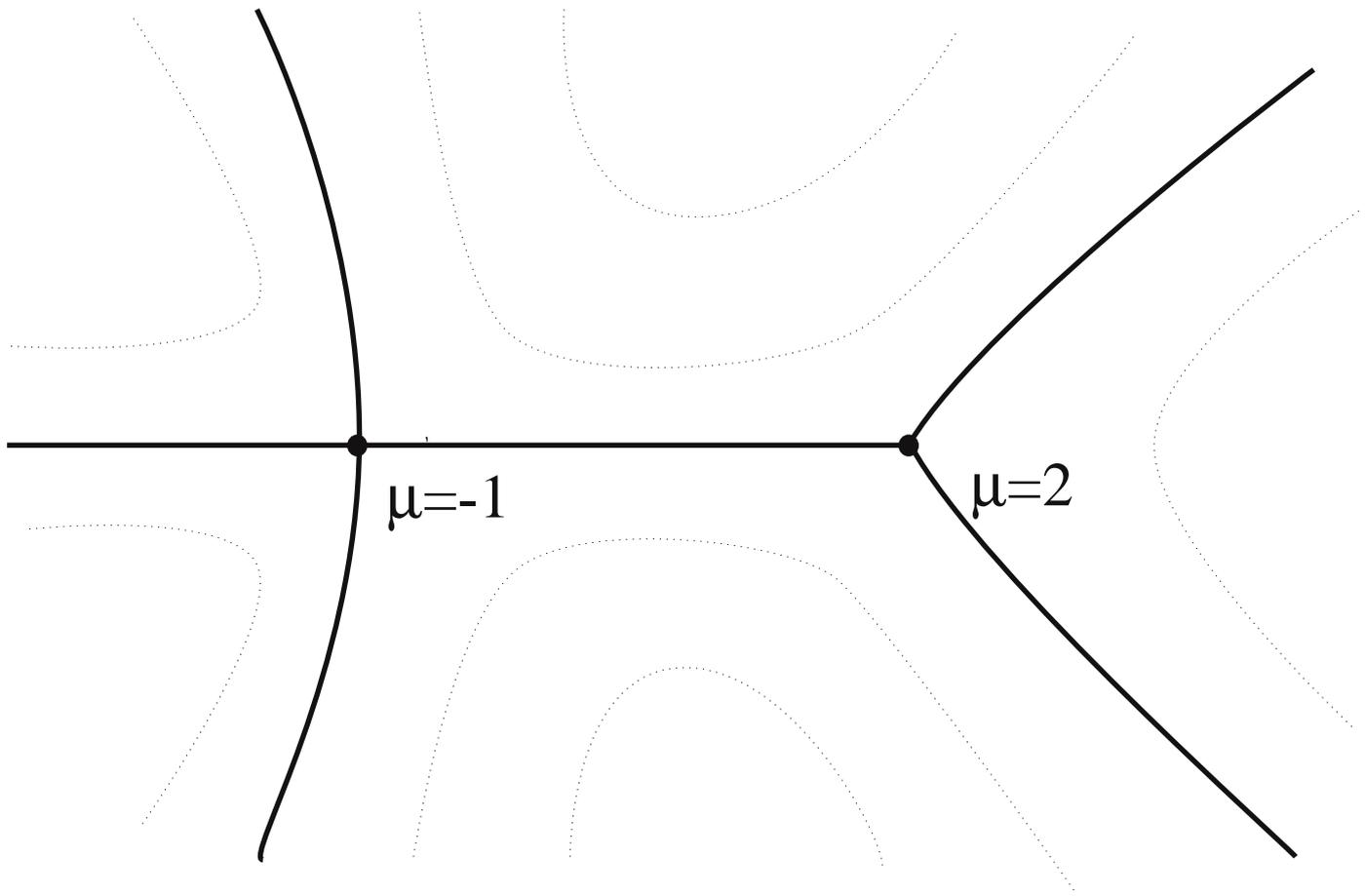
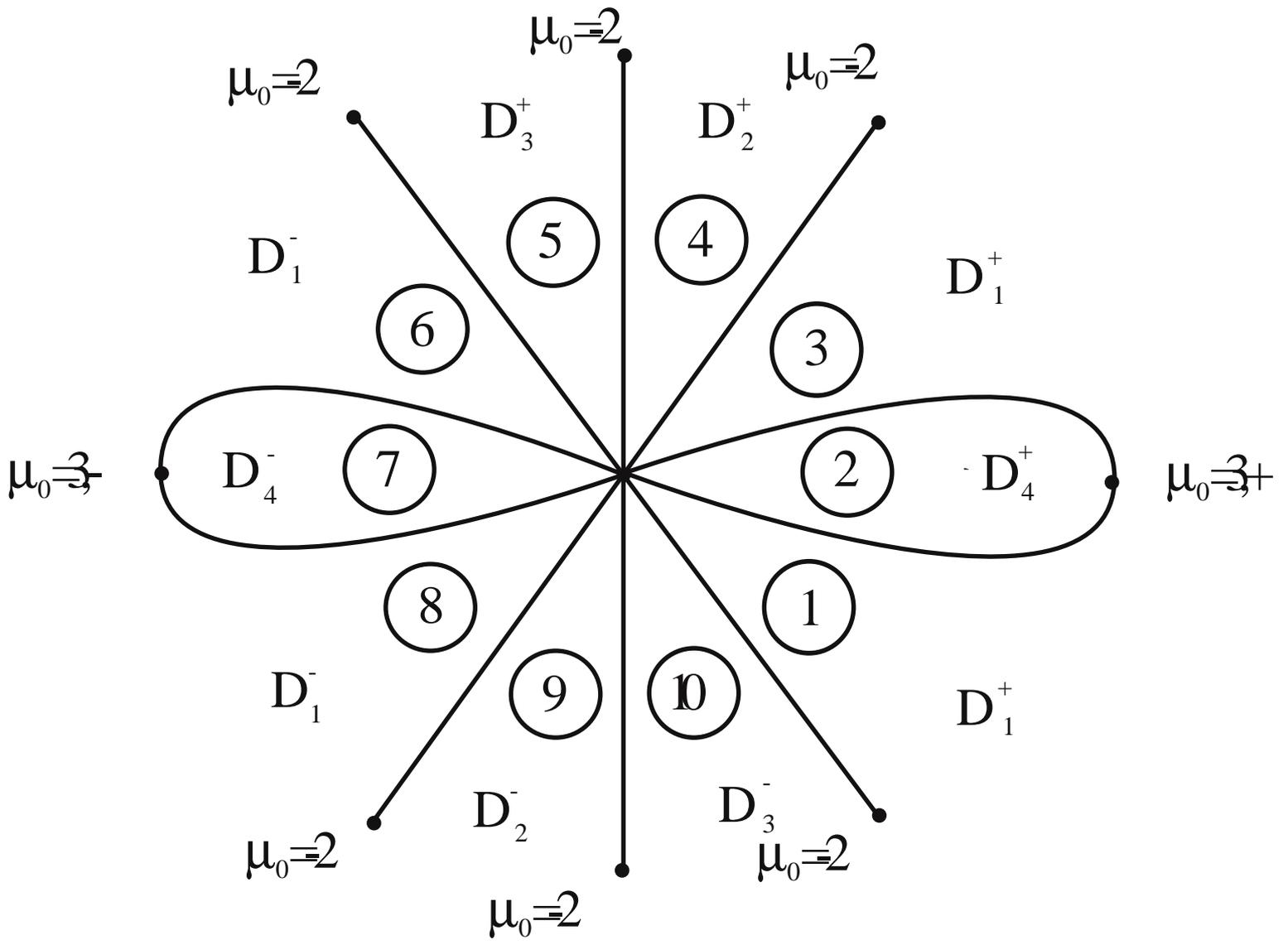
\end{document}